\documentclass[a4paper,10pt,oneside,twocolumn]{article}
\usepackage{graphicx}
\usepackage[T1]{fontenc}
\usepackage{mathptmx}
\usepackage[scaled=.92]{helvet}
\usepackage{courier}
\usepackage{fancyhdr}
\usepackage{authblk}
\usepackage[original]{abstract}

\graphicspath{{./}{figures/}}

\usepackage[utf8]{inputenc}

\usepackage{microtype}
\usepackage{mathtools}
\usepackage{amsfonts}
\usepackage{graphicx}
\usepackage{gensymb}

\usepackage[numbers,square]{natbib}

\usepackage{booktabs}
\usepackage{color}
\usepackage{url}
\usepackage{multirow}
\usepackage{hyperref}

\hypersetup{
  linkcolor  = black,
  citecolor  = black,
  urlcolor   = black,
  colorlinks = true,
}

\makeatletter
 \DeclareRobustCommand\ref{%
    \@ifstar\@refstar\T@ref
  }%
  \DeclareRobustCommand\pageref{%
    \@ifstar\@pagerefstar\T@pageref
  }%
 \makeatother

\DeclareMathOperator{\diag}{diag}

\newcommand{\tcircle}[1]{\textcircled{\raisebox{-0.9pt}{#1}}}

\title{Time-of-arrival estimation and phase unwrapping of head-related transfer functions with integer linear programming}

\author{Chin-Yun Yu}
\author{Johan Pauwels}
\author{Gy{\"o}rgy Fazekas}

\affil{Centre for Digital Music, Queen Mary University of London}

\date{\hyperlink{mailto:chin-yun.yu@qmul.ac.uk}{chin-yun.yu@qmul.ac.uk}}

\begin{document}
\pagestyle{fancy}

\twocolumn[
\maketitle 

\begin{onecolabstract}
In binaural audio synthesis, aligning head-related impulse responses (HRIRs) in time has been an important pre-processing step, enabling accurate spatial interpolation and efficient data compression.
The maximum correlation time delay between spatially nearby HRIRs has previously been used to get accurate and smooth alignment by solving a matrix equation in which the solution has the minimum Euclidean distance to the time delay. 
However, the Euclidean criterion could lead to an over-smoothing solution in practice. 
In this paper, we solve the smoothing issue by formulating the task as solving an integer linear programming problem equivalent to minimising an $L^1$-norm. 
Moreover, we incorporate 1) the cross-correlation of inter-aural HRIRs, and 2) HRIRs with their minimum-phase responses to have more reference measurements for optimisation. 
We show the proposed method can get more accurate alignments than the Euclidean-based method by comparing the spectral reconstruction loss of time-aligned HRIRs using spherical harmonics representation on seven HRIRs consisting of human and dummy heads.
The extra correlation features and the $L^1$-norm are also beneficial in extremely noisy conditions.
In addition, this method can be applied to phase unwrapping of head-related transfer functions, where the unwrapped phase could be a compact feature for downstream tasks.
\end{onecolabstract}
]

\section{Introduction}

Head-related transfer functions (HRTFs) describe the response of the travelling path from the sound source to human ears in different directions.
These functions encode the acoustic scattering caused by the head, pinna, and torso, which humans have learnt to interpret as cues to perceive the spatial direction of the sound. 
Differences in human morphology make HRTFs differ between individuals.
Using personalised HRTFs to simulate virtual acoustic environments creates an immersive user experience in games, virtual reality (VR) applications and the metaverse.

Theoretically, HRTFs vary continuously when the sound sources move from one direction to another.
Continuous representation is feasible on simple objects, e.g., a rigid sphere model, but not for human subjects with complex shapes.
The common practice is to measure HRTFs in an anechoic chamber on discrete spatial directions and use interpolation methods to fill the gap between directions, assuming the spatial resolution is dense enough.
Various interpolation methods have been proposed, including linear interpolation~\citep{gamper2013selection}, principle component analysis~\citep{wang_head-related_2009}, and interpolation of spherical harmonics (SHs)~\citep{evans1998analyzing, zaar2011phase, brinkmann2018comparison, duraiswami_interpolation_nodate, arend_spatial_nodate, ben-hur_efficient_2019, reijniers_noise-resistant_2023, arend2023mag}.
We focus on SHs interpolation in this paper.

The choice of representation affects interpolation accuracy and the required computational resources.
Using HRIRs directly requires many SHs to reduce interpolation errors~\citep{zaar2011phase}.
A lot of the spatial complexity comes from the differences in the travelling time of sound in each direction, known as time-of-arrivals (TOAs).
Various representations have been used to mitigate this effect, including time-aligned HRIRs~\citep{reijniers_noise-resistant_2023, arend2023mag}, calibrating ear positions to the centre of the head~\citep{ben-hur_efficient_2019}, or inversely filtering HRTFs with a fixed rigid sphere HRTFs~\citep{arend_spatial_nodate}.
For time alignment, using the correlation between paired HRIRs to measure time differences has shown to be highly resistant to noise~\citep{reijniers_noise-resistant_2023, porschmann_cross-correlation-based_2023}.

Naturally, many different HRIR pairings can be made, and due to the spherical sampling, multiple pairs can be equidistant. 
Therefore, the time alignment is best done globally instead of pairwise.
Given the measured time differences, \citet{reijniers_noise-resistant_2023} consider all the directions at once and formalise the TOA estimation problem as solving a matrix equation, where the derivatives to TOAs of the Euclidean distance to the measurements are zeros, achieving state-of-the-art (SOTA) in HRIRs alignment.
However, least square solutions are known to be over-smooth in practice and more sensitive to outliers from the measurement (here, we mean the noisy time differences) compared to minimising $L^1$-norm~\citep{pml1Book}.

This paper aims to solve the over-smoothing and noise-sensitivity issues by extending the SOTA to minimise $L^1$-norm with integer linear programming (ILP).
We connect HRTF measurement directions into a graph with edges representing the measured difference in the target quantity.
We formalise the problem as minimising residuals on edges, subjecting to the constraint that the differences in every elementary cycle must sum to zero.
This method strongly relates to the minimum-cost network flow approach in the task of phase unwrapping (PU)~\citep{costantini_novel_1998} and, to the best of the authors' knowledge, has not been applied to HRIRs TOA estimation nor even HRTFs PU, where the latter combining with HRTFs magnitude responses was shown to be a suitable representation for interpolation as well~\citep{evans1998analyzing, zaar2011phase, brinkmann2018comparison}.
Moreover, we propose adding two other correlation-based time difference features as extra hints for TOA estimation.

The paper is organised as follows.
Section~\ref{sec:tos} and~\ref{sec:pu} detailed our proposed TOA estimation and PU methods.
Section~\ref{sec:eval} shows our evaluations in comparing the reconstructed errors of time-aligned HRIRs on seven different HRTFs.
We then compare the robustness of our methods to~\citet{reijniers_noise-resistant_2023} with simulated noisy HRTFs.
Finally, we show the unwrapped phase of the SONICOM KEMAR HRTFs~\citep{engel_sonicom_2023} using the proposed method as a demonstration.

\section{HRIRs time-of-arrival estimation}
\label{sec:tos}

Given a subject's HRIRs measured in $N$ directions on a 2-sphere $\theta_i \in \mathbb{S}^2$, we wish to calculate the TOAs $\tau_i^{L/R}$ of each HRIR $h_i^{L/R}[t]$ in $\theta_i$.
In the following discussions, we treat each ear independently for brevity and use the left ear as an example.
Unless we mention it specifically, we treat $\tau_i^L$ as $\tau_i$.

Assuming we do not know $\tau_i$ but we know the time differences between arbitrary directions $\theta_i$ and $\theta_j$: $\Gamma_{i,j} = \tau_j - \tau_i ~\forall (i,j)$. We then create a connected graph $G = (V(G), E(G))$ by considering the directions as a set of vertices $V(G) = \{0, 1,\dots,N-1\}$. Its edges $E(G)$ are drawn for every pair of vertices for which the time difference $\Gamma_{i,j}$ is known. Any TOA $\tau_i$ can then be obtained, apart from a constant, by integrating time differences $\Gamma_e$ along the path from a reference vertex to vertex $i$.
\begin{equation}
    \tau_i = \sum_{e \in \Phi_G(0, i)} \Gamma_e + C,
\label{eq:integral}
\end{equation}
in which $\Phi_G(0, i) \subset E(G)$ are edges that form a path from vertex 0 to $i$.
We chose 0 as the starting point for convenience.
The unknown $C$ is not an issue when calculating inter-aural time differences (ITDs, $\tau_i^L - \tau_i^R $) and aligning HRIRs, which we will discuss later in Sec.~\ref{ssec:lstsq}.

\subsection{Cross-correlation-based estimation}

We describe $\Gamma_{i,j}$ with $\hat{\Gamma}_{i,j} = \arg\max_t h_i[t] \star h_j[t]$ and a residual $K_{i,j}$ as
\begin{equation}
    \Gamma_{i,j} = \hat{\Gamma}_{i,j} + K_{i,j}.
\label{eq:gamma_+k}
\end{equation}
If the correlation-based method is accurate enough, $|K_{i,j}|$ should be small.
Estimating $\Gamma_{i,j}$ becomes minimising $|K_{i,j}|$ while being subject to constraints $\mathtt{C}(K; \hat{\Gamma}, G)$ that we chose.

\subsection{The irrotational property of $\Gamma$}

Any cycle $P \subset G$ represents a path with the same vertex as the start and end points.
Plugging $P$ into Eq.~\eqref{eq:integral} results in
\begin{equation}
    \sum_{(u, v) \in E(P)}\Gamma_{u, v} = \sum_{v \in V(P)}\tau_v - \sum_{u \in V(P)}\tau_u = 0. 
\label{eq:irrotational}
\end{equation}
One way to think of it is that $\tau(\theta)$ is a continuous function differentiable everywhere on $\mathbb{S}^2$, and for any enclosed path $P$ on $\mathbb{S}^2$, the integration $\int_{u\in P} \nabla \tau(u) \partial u = 0$.
The same applies to discrete $\tau_i$.
We utilise this irrotational property and substitute Eq. \eqref{eq:gamma_+k} into \eqref{eq:irrotational} to get
\begin{equation}
\sum_{e \in E(P)}K_{e} = - \sum_{d \in E(P)}\hat{\Gamma}_{d}.
\label{eq:single_constraint}
\end{equation}

Multiple solutions of $K_e$ to Eq. 4 exist with just one cycle.
Assuming there are cycles $S = \{P^0, P^1, \dots\}$ where $\cup_{P \in S} E(P) \subseteq E(G) = \{e^0, e^1,\dots\}$, we can extend Eq.~\eqref{eq:single_constraint} into the following matrix equation:
\begin{equation}
    \begin{gathered}
    \mathbf{A} \mathbf{k} = - \mathbf{A} \mathbf{x} \\
    \mathbf{x} = [\hat{\Gamma}_{e^0}, \hat{\Gamma}_{e^1}, \dots]^T\\
    \mathbf{k} = [K_{e^0}, K_{e^1},\dots]^T\\
    A_{ij} =
        \begin{rcases}
            \begin{dcases}
                -1, & (v, u) \in E(P^i) \\
                1, & (u, v) \in E(P^i) \\
                0, & otherwise
            \end{dcases}
        \end{rcases}: (u, v) = e^j.
    \end{gathered}
\label{eq:contraint}
\end{equation}
We use Eq. \eqref{eq:contraint} as our constraint $\mathtt{C}(K; \hat{\Gamma}, G)$.
Because the unit of $\hat{\Gamma}$ is the number of audio samples, the solution to $\Gamma$ are integers.
Putting this all together, estimating the time difference $\Gamma$ becomes solving the following ILP problem:
\begin{equation}
    \begin{gathered}
        \min\limits_\mathbf{k} ~\mathbf{w}^T|\mathbf{k}|\\
        s.t. ~\mathbf{A} \mathbf{k} = - \mathbf{A} \mathbf{x},
    \end{gathered}
\label{eq:ilp_cycles}
\end{equation}
where $\mathbf{w} = [w_{e^0}, w_{e^1},\dots]^T$ is a weighting vector.

\subsection{The cycle-less approach}
The size of $\mathbf{A}$ is $|S| \times |E(G)|$, and solving Eq.~\eqref{eq:ilp_cycles} can be computationally intensive if the number of cycles $|S|$ is large.
Moreover, if $S$ cannot be easily defined when constructing $G$, searching the necessary cycles in $G$ can be time-consuming.
We can convert the constraint from the cycles to the edges according to the edgelist method~\citep{shanker_edgelist_2010, ma_time_2022}, resulting in

\begin{equation}
    \begin{gathered}
        \min\limits_{\mathbf{k}, \mathbf{\tau}} ~\mathbf{w}^T|\mathbf{k}|\\
        s.t. ~\begin{bmatrix}
            \mathbf{A}^{edge} & \mathbf{I}
        \end{bmatrix}
        \begin{bmatrix}
            \mathbf{\tau} \\
            \mathbf{k}
        \end{bmatrix} = \mathbf{x} \\
        A_{ij}^{edge} = 
        \begin{rcases}
            \begin{dcases}
                -1, & j = u \\
                1, & j = v\\
                0, & otherwise
            \end{dcases}
        \end{rcases}: (u, v) = e^i.
    \end{gathered}
\label{eq:ilp_edgelist}
\end{equation}
This formulation solves $\tau$ jointly and does not require integration (Eq.~\eqref{eq:integral}) afterwards.
The size of the matrix $[\mathbf{A}^{edge}~\mathbf{I}]$ is $|E(G)| \times (|E(G)| + N)$.

The two formulations are equal as long as $\cup_{P \in S} E(P) = E(G)$~\cite{shanker_edgelist_2010}.
Nevertheless, the required time to solve them varies depending on the structure of $G$ and the selected $S$.

\subsection{The least square approach}
\label{ssec:lstsq}
Prior work by~\citet{reijniers_noise-resistant_2023} minimises the following criterion:
\begin{equation}
    \begin{split}
        \min\limits_{\tau} \sum_{(u, v) \in E(G)} w_{u, v} [\hat{\Gamma}_{u, v} - (\tau_v - \tau_u)]^2 \\
        + \lambda \sum_{u \in V(G)} \tau_u.
    \end{split}
\label{eq:l2_criterion}
\end{equation}
The second term is for regularisation purposes.
If we neglect the regularisation term, plug Eq. \eqref{eq:gamma_+k} into Eq. \eqref{eq:l2_criterion}, it is the same as minimising Eq. \eqref{eq:ilp_edgelist} but the objective becomes quadratic: $\min\limits_{\mathbf{k}, \mathbf{\tau}}~\mathbf{k}^T \diag(\mathbf{w}) \mathbf{k}$.
In other words, we are optimising the same equation from~\cite{reijniers_noise-resistant_2023} but using $L^1$-norm as the criterion, not $L^2$.

The optimal solution is when the derivatives of Eq. \eqref{eq:l2_criterion} to $\tau$ and $\lambda$ are zeros, which leads to solving the following equation:
\begin{equation}
    \begin{gathered}
        \begin{split}
        \begin{bmatrix}
             \diag(\mathbf{W} \cdot \mathbf{1}) - \mathbf{W}\\
             \mathbf{1}
        \end{bmatrix} \mathbf{\tau} \\
        = \begin{bmatrix}
            \diag(\mathbf{W}\mathbf{X}) + \lambda & 0
        \end{bmatrix} 
        \end{split}\\
        W_{ij} =
        \begin{rcases}
            \begin{dcases}
                w_{i,j}, & (i,j) \in E(G) \\
                w_{j,i}, & (j,i) \in E(G) \\
                0, & otherwise
            \end{dcases}
        \end{rcases} \\
        X_{ij} = 
        \begin{rcases}
            \begin{dcases}
                \hat{\Gamma}_{i,j}, & (i,j) \in E(G) \\
                -\hat{\Gamma}_{i,j}, & (j,i) \in E(G) \\
                0, & otherwise
            \end{dcases}
        \end{rcases}.
    \end{gathered}
\label{eq:l2_lstsq}
\end{equation}
The matrix size of the equation is $(|V(G)|+1) \times |V(G)|$.

We set $C$ to the same value for both ears so it can be cancelled after the ITD subtraction.
We did this by assuming $\sum_{i=0}^{N-1} \tau_i^L = \sum_{i=0}^{N-1} \tau_i^R = 0$, which is a good approximation due to the symmetric characteristic of HRIRs.
For HRIRs alignment, we use $\overset{\star}{\tau_i} = \tau_i - \min \tau_i$.

\subsection{Inter-aural cross-correlation features}
So far, we treat each ear independently without considering any inter-aural relation.
The inter-aural cross-correlation $h_i^L[t] \star h_i^R[t]$ has been used to calculate ITDs before~\cite{ kistler1992model}.
This feature can be added easily to all the methods above by adding $N$ extra edges to connect the two graphs for each ear.
Assuming the graph for the left ear $G^L = G$ and the right ear $G^R = (\{N,N+1,\dots,2N-1\}, \{(i+N,j+N): (i,j) \in E(G)\})$, the joint graph is $G^{joint} = (\{0, 1,\dots,2N-1\}, E(G^L) \cup E(G^R) \cup \{(i,i+N): 0 \leq i < N\})$.

We estimate the inter-aural difference simply as $\hat{\Gamma}_{i,i+N} = \arg\max_t h_i^L[t] \star h_i^R[t]$.
The assumption of $C$ being equal for both ears is unnecessary as the inter-aural relation is jointly estimated.

\subsection{Incorporating absolute time information}
The maximum correlation of $h_i[t]$ with its minimum-phase version $h_i^{min}[t]$ is a close estimate of $\tau_i$ based on the assumption that $h_i[t] \approx h_i^{min}[t - \tau_i]$, which holds in most directions~\cite{nam2008method}.
We can also incorporate this into the above methods.
Let us introduce an auxiliary position $\delta: \tau_\delta = 0$ and define $\hat{\Gamma}_{\delta,i} = \arg\max_t h_i^{min}[t] \star h_i[t]$.
Eq.~\eqref{eq:ilp_cycles} can be applied straightforwardly with $G^{\delta} = (\{\delta\} \cup V(Q), \{(\delta, i): 0 \leq i \leq \max V(Q)\} \cup E(Q)\})$ where $Q \in \{G, G^{joint}\}$.
Note that we can use integration path $\Phi_{G^\delta}(\delta, i)$ to cancel $C$ because we know $\tau_\delta = 0$.

For the cycle-less approach, replace $\mathbf{A}^{edge}$ with $\mathbf{A}^\delta$, where
\begin{equation}
    A_{ij}^{\delta} = 
    \begin{rcases}
        \begin{dcases}
            -1, & j = u \cap u \neq \delta \\
            1, & j = v\\
            0, & otherwise
        \end{dcases}
    \end{rcases}: (u, v) = e^i.
\label{eq:A_delta}
\end{equation}

\begin{figure*}[t]
\begin{center}
\includegraphics[width=0.97\textwidth]{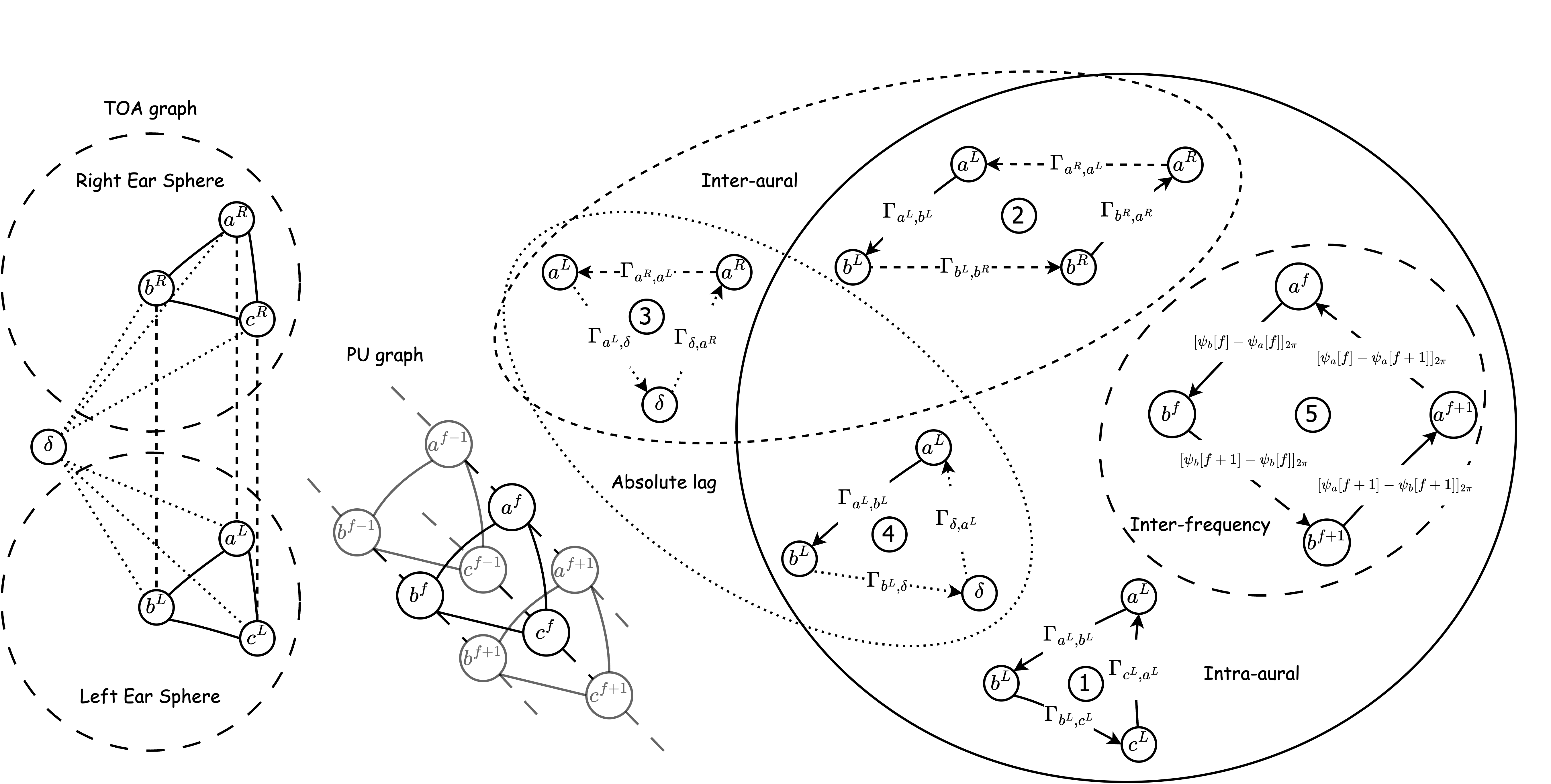}
\caption{Two simple graphs consists of three measurement directions $\{a,b,c\}$, and an auxiliary vertex $\delta$ with $\tau_\delta=0$. The elementary cycles in graphs can be categorised into five groups on the right, with four types of edges: inter/intra-aural time differences, absolute time lags, and inter-frequency phase differences. We use (\tcircle{1}-\tcircle{4}) for TOA estimation and (\tcircle{1}, \tcircle{5}) for PU.}
\label{fig:graph_example}
\end{center}
\end{figure*}

We propose a similar version of Eq.~\eqref{eq:l2_lstsq} with $G^\delta$ as
\begin{equation}
    \begin{gathered}
        \begin{split}
            \diag(\mathbf{W} \cdot \mathbf{1} + \mathbf{w}^\delta) \mathbf{\tau} - \mathbf{W} \mathbf{\tau} \\
            = \diag(\mathbf{W}\mathbf{X}) + \diag(\mathbf{w}^\delta) \mathbf{x}^\delta
        \end{split}\\
        \mathbf{x}^\delta = [\hat{\Gamma}_{\delta, 0}, \hat{\Gamma}_{\delta, 1},\dots]^T\\
        \mathbf{w}^\delta = [w_{\delta, 0}, w_{\delta, 1},\dots]^T
    \end{gathered}
\label{l2:eq_solve}
\end{equation}
which is when the derivatives with respect to $\tau$ of the following equation are zeros:
\begin{equation}
    \begin{split}
        \min\limits_{\tau} \sum_{(u, v) \in E(G^\delta)} 
        w_{u, v} \left[\hat{\Gamma}_{u, v} - (\tau_v - \tau_u)\right]^2.
    \end{split}
\label{eq:l2_criterion_delta}
\end{equation}
The size of the matrix equation is $(|V(G^\delta)| - 1)^2$. 
An example of $G^\delta$ is shown in Fig.~\ref{fig:graph_example}.

\subsection{The choice of weighting $\mathbf{w}$}
\label{ssec:weighting}

Exponential weighting based on the distance between directions has been proposed in~\cite{reijniers_noise-resistant_2023}:

\begin{equation}
\label{eq:weight_intra}
    w_{i,j} = \exp\left(-\frac{\cos^{-1} (\theta_i \cdot \theta_j)}{\sigma}\right),
\end{equation}
with $\sigma$ set to $8^{\degree}$ perform the best.
This weighting emphasises finer local details, which is reasonable because HRIRs are more similar for closer directions, and their correlations are more accurate.
For inter-aural edges $w_{i, i + N}$, we replace $\theta_j$ in Eq.~\eqref{eq:weight_intra} to $\theta_i$'s reflection through the plane $y = 0$, like virtually placing the other ear on the symmetric position.
For $w_{\delta, i}$, we sampled a few values in $[0.1, 1]$ and empirically found 0.1 works well, therefore we use 0.1 in the rest of the paper.

The exponential weighting requires two hyperparameters $\sigma$ and a constant weight for $w_{\delta, i}$.
To counter this, we propose a parameter-free weighting method, which is setting $w_{i,j} = \sum_{t} \Tilde{h}_i[t-\hat{\Gamma}_{i,j}] \Tilde{h}_j[t]$ where $\Tilde{h}[t]$ is the normalised $h[t]$ so $0 \leq w_{i,j} \leq 1$ (same for $w_{\delta, i} = \sum_{t} \Tilde{h}_i^{min}[t-\hat{\Gamma}_{\delta,i}] \Tilde{h}_i[t]$).
This is based on our hypothesis that if two HRIRs have similar waveforms, then their maximum correlation should be high, and their $\hat{\Gamma}_{i,j}$ is closer to $\Gamma_{i,j}$.

\section{HRTFs phase unwrapping}
\label{sec:pu}
The phase is a \emph{temporal} quantity representing periodic signals' angular position.
In HRTFs, it is a continuous function of spatial coordinate and angular frequency $\phi(\theta, \omega)$.
We can measure it from HRTFs $H_i[f] \in \mathbb{C}$ at frequency index $f$, which defines discrete sample points over $\omega$.
However, although the $\phi_i[f]$ can be any real number, the measured phase $\psi_i[f] = \tan^{-1}\left(Im(H_i[f])/Re(H_i[f])\right)$ is wrapped inside $[-\pi, \pi)$.
Their relation is represented as
\begin{equation}
    \phi_i[f] = \psi_i[f] + 2\pi L_i[f], L_i[f] \in \mathbb{Z}.
\end{equation}
The task of HRTFs PU is to find a set of residual $L_i[f]$ that best describes the phase response.

\subsection{Unwrapping along the frequency axis}
\label{ssec:freq-pu}

The simplest solution~\cite{evans1998analyzing} is integrating the wrapped phase differences along the frequency axis:
\begin{equation}
    \phi_i[f] = \phi_i[0] + \sum_{l=1}^{f} \left[\psi_i[l] - \psi_i[l-1]\right]_{2\pi},
\end{equation}
where $\phi_i[0] = \psi_i[0]$ and $[x]_{2\pi} = x + \pi \pmod{2\pi} - \pi$.
This method is only valid when the Itoh condition~\citep{Itoh:82} is met, which is $|\phi_i[f] - \phi_i[f-1]| \leq \pi$ so the wrapped difference equals to the true difference.
Violating this condition creates an aliasing effect and results in ambiguous phase differences.

\subsection{Unwrapping across $\mathbb{S}^2$}
\label{ssec:sph-pu}
\citet{zaar2011phase} is the first and only work that uses phase differences between the nearby spatial directions for PU.
In their method, PU is performed on each frequency independently.
Starting from the ipsilateral side of the sphere, where the HRIR has the strongest energies and less time delay, a set of rules is used to unwrap the directions along the way towards the contralateral side.

This method creates significantly fewer spatial discontinuities across the sphere than the naive frequency-based method.
However, as the frequency increases, the Itoh condition is less likely to be met, as more periods are needed for high frequencies to travel the same given distance.
Phase differences along the frequency axis are more reliable and could therefore enhance the performance of spherical unwrapping.

\subsection{Unwrapping with both $\mathbb{S}^2$ and $f$ axis}
\label{ssec:sph-freq-pu}

Given the previous explanation, PU equals TOA estimation with an extra frequency dimension.
As long as the wrapped phase can be represented as a graph, we can plug it directly into Eq.~\eqref{eq:ilp_cycles} or~\eqref{eq:ilp_edgelist} to calculate $L_i[f]$.
Only the definition of $\hat{\Gamma}$ is changed, where we use the wrapped phase differences to substitute time differences.

We view the frequency index $f$ in each direction $\theta_i$ as a single vertex.
$F$ is the number of frequency bins.
We use the same graph $G$ from Section~\ref{sec:tos} for each frequency, defined as $\{G^f: 0 \leq f < F\}$ and $G^f = (\{fN, fN+1,\dots,(f+1)N - 1\}, \{(i + fN, j + fN): (i,j) \in E(G) \})$.
We add edges along the frequency axis $E^\psi = \{(v, v + N): v \in \{0, 1,\dots, (F-1)N - 1\}$ to connect $\psi_i[f]$ and $\psi_i[f+1]$, resulting in one joint graph $G^\psi = (\{0, 1,\dots, FN - 1\}, \cup_{0 \leq f < F} E(G^f) \cup E^\psi \})$.
An example of $G^\psi$ is shown in Fig.~\ref{fig:graph_example}.

To simplify notation, we use $\phi_u$ to represent $\phi_{u\pmod{N}}[\lfloor \frac{u}{N} \rfloor]$ and use the same for similar variables.
The normalised phase differences are
\begin{equation}
\label{eq:phase_diff}
    \begin{split}
        \frac{\phi_v - \phi_u}{2\pi} &= \frac{1}{2\pi} \left(\psi_v - \psi_u + 2\pi(L_v - L_u)\right) \\
                                     &= \frac{[\psi_v - \psi_u]_{2\pi}}{2\pi} + K_{u,v}.
    \end{split}
\end{equation}
The RHS of Eq.~\eqref{eq:phase_diff} becomes Eq.~\eqref{eq:gamma_+k}'s RHS when we substitute $\frac{\left[\psi_v - \psi_u\right]_{2\pi}}{2\pi}$ with $\hat{\Gamma}_{u,v}$.
If we further assume that the residuals $K_{u,v}$ are close to zeros in the HRTFs, solving it becomes the same problem as TOA estimation.
It is a widely used method for PU on synthetic aperture radar data~\citep{costantini_novel_1998, shanker_edgelist_2010, ma_time_2022}.

To solve the violations of the Itoh condition (high amount of nonzero $K_{u,v}$) in high frequencies, we propose unwrapping the phase on the aligned HRIRs instead.
This trick greatly reduces the spherical phase differences between directions.

\section{Evaluations and discussions}
\label{sec:eval}

\begin{table*}[t]
\caption{Details of HRTFs used for evaluation, consisting of dummy heads and real subjects. The runtime was measured on a MacBook with M1 Pro and averaged across all 12 configurations for each method.}
\label{tab:datasets}
\centering
\resizebox{1\textwidth}{!}{%
\begin{tabular}{@{}lllrrccc@{}}
    \toprule
    \multirow{2}{*}{\textbf{Database}} & 
    \multirow{2}{*}{\textbf{Simplified filename (.sofa)}} & 
    \multirow{2}{*}{\textbf{fs (kHz)}} & 
    \multicolumn{1}{l}{\multirow{2}{*}{\textbf{N}}} & 
    \multicolumn{1}{l}{\multirow{2}{*}{$\left|\mathbf{E(G)}\right|$}} &
    \multicolumn{3}{c}{\textbf{TOAs runtime (s)}} 
    \\ \cline{6-8}
    &&&&& \textbf{LS} & \textbf{SIMP} & \textbf{EDGY}
    \\
    \midrule
    AACHEN & MRT01                       & 44.1 & 2304 & 9141 & 0.164 & 1.041 & 1.652  \\
    ARI    & hrtf\_nh4                   & 48   & 1550 & 6119 & 0.072 & 0.540 & 0.479 \\
    CIPIC  & subject\_003                & 44.1 & 1250 & 4963 & 0.055 & 0.414 & 0.341 \\
    RIEC   & *\_subject\_001             & 48   & 865  & 3385 & 0.033 & 0.282 & 0.265 \\
    SADIE  & H3\_*\_256tap\_FIR          & 48   & 2818 & 11266 & 0.120 & 3.559 & 2.668 \\
    SONICOM & KEMAR\_*\_SmallEars\_FreeFieldComp\_* & 44.1 & 828 & 3237 & 0.042 & 0.193 & 0.166 \\
    THK    & HRIR\_L2702\_NF100          & 48   & 2702 & 10802 & 0.098 & 1.609 & 1.317 \\ 
    \bottomrule
\end{tabular}%
}
\end{table*}

\begin{table}[]
\centering
\caption{The best configuration for each dataset to achieve the lowest LSD.}
\label{tab:best_config}
\resizebox{0.95\columnwidth}{!}{%
\begin{tabular}{cccccc}
\toprule
\textbf{Dataset} & \textbf{Alg.} & \textbf{Min.} & \textbf{Cross} & \textbf{Weight} & \textbf{LSD (dB)} \\ \midrule
AACH.   & SIMP & w/o  & w/o  & NONE  & 3.61 \\ \midrule
ARI     & SIMP & w/   & w/o  & EXP   & 3.82 \\ \midrule
CIPIC   & LS   & w/o  & w/o  & NONE  & 3.08 \\ \midrule
RIEC    & EDGY & w/   & w/   & EXP   & 3.19  \\ \midrule
SADIE   & SIMP & w/o  & w/o  & NONE  & 5.01 \\ \midrule
SONIC.  & EDGY & w/o  & w/o  & NONE  & 2.68  \\ \midrule
THK     & EDGY & w/o  & w/o  & CORR  & 3.01 \\ 
\bottomrule
\end{tabular}%
}
\end{table}

\begin{table*}[h!]
\centering
\caption{The average ITD distortion ($\mu$s) and HRTFs LSD (dB) across different configurations.}
\label{tab:method_compare}
\resizebox{1\textwidth}{!}{%
\begin{tabular}{ccccccccccccccc}
\toprule
 &
    \multicolumn{2}{c}{\textbf{AACHEN}} &
    \multicolumn{2}{c}{\textbf{ARI}} &
    \multicolumn{2}{c}{\textbf{CIPIC}} &
    \multicolumn{2}{c}{\textbf{RIEC}} &
    \multicolumn{2}{c}{\textbf{SADIE}} &
    \multicolumn{2}{c}{\textbf{SONICOM}} &
    \multicolumn{2}{c}{\textbf{THK}} \\ 
    \midrule
    \multicolumn{1}{c}{\textbf{Algorithm}}&
    \textbf{$\triangle$ITD} & \textbf{LSD} &
    \textbf{$\triangle$ITD} & \textbf{LSD} &
    \textbf{$\triangle$ITD} & \textbf{LSD} &
    \textbf{$\triangle$ITD} & \textbf{LSD} &
    \textbf{$\triangle$ITD} & \textbf{LSD} &
    \textbf{$\triangle$ITD} & \textbf{LSD} &
    \textbf{$\triangle$ITD} & \textbf{LSD} \\ 
    \midrule
    EDGY &
    12.26 & \textbf{3.80} &
    17.68 & \textbf{4.08} &
    13.06 & \textbf{3.34} &
    13.22 & \textbf{3.35} &
    \textbf{31.22} & \textbf{5.45} &
    8.41 & \textbf{2.82} &
    8.51 & \textbf{3.30} 
    \\ 

    SIMP &
    12.30 & \textbf{3.80} &
    17.70 & \textbf{4.08} &
    13.09 & \textbf{3.34} &
    13.26 & \textbf{3.35} &
    31.27 & 5.46 &
    8.48 & \textbf{2.82} &
    8.54 & \textbf{3.30} 
    \\
    LS &
    \textbf{11.32} & 3.96 &
    \textbf{17.22} & 4.20 &
    \textbf{12.42} & 3.51 &
    \textbf{12.65} & 3.46 &
    31.44 & 6.30 &
    \textbf{7.81} & 2.98 &
    \textbf{8.40} & 3.65
    \\ 
    \midrule
    \multicolumn{1}{c}{\textbf{Weight}}\\ \midrule 
    EXP &
    \textbf{11.00} & \textbf{3.73} &
    \textbf{17.32} & \textbf{3.90} &
    \textbf{12.77} & \textbf{3.26} &
    \textbf{12.32} & \textbf{3.26} &
    \textbf{28.24} & \textbf{5.63} &
    \textbf{7.15} & \textbf{2.75} &
    \textbf{6.87} & \textbf{3.23}\\
    
    CORR & 
    12.32 & 3.91 &
    17.60 & 4.22 &
    12.89 & 3.45 &
    13.26 & 3.44 &
    31.72 & 5.78 &
    8.67 & 2.92 &
    9.18 & 3.51\\
    
    NONE &
    12.55 &  3.92 &
    17.67 &  4.25 &
    12.91 &  3.47 &
    13.54 &  3.47 &
    33.96 &  5.80 &
    8.88 &  2.95 &
    9.40 &  3.51 \\ 
    \bottomrule
\end{tabular}%
}
\end{table*}

We followed previous works~\citep{gamper2013selection, zaar2011phase} to build a graph from the measurements using the convex hull algorithm.
We set $\lambda$ in Eq.~\eqref{eq:l2_criterion} to 0.1.
All the HRIRs were oversampled ten times before computing correlations for a finer time resolution.
We set all the weights to one for the PU experiment.
We use the regularised least square method~\cite{duraiswami_interpolation_nodate} with a regularisation of $10^{-5}$ to perform SH transform.
Table~\ref{tab:datasets} summarises the HRTFs we used in this paper.

We tested three algorithms: 
\vspace{-2mm}
\begin{itemize}
    \setlength\itemsep{-0.3em}
    \item \textbf{SIMP}: $L^1$ criterion, simplices (cycles)-based (Eq.~\eqref{eq:ilp_cycles})
    \item \textbf{EDGY}: $L^1$ criterion, edgelist-based (Eq.~\eqref{eq:ilp_edgelist})
    \item \textbf{LS}: $L^2$ criterion, least squared solutions (Eq.~\eqref{eq:l2_lstsq} and~\eqref{eq:l2_criterion_delta})
\end{itemize}
\vspace{-2mm}
and three weighting schemes:
\vspace{-2mm}
\begin{itemize}
    \setlength\itemsep{-0.3em}
    \item \textbf{NONE}: uniform weights
    \item \textbf{EXP}: exponential weights~\cite{reijniers_noise-resistant_2023}
    \item \textbf{CORR}: proposed correlation weights in Sec.~\ref{ssec:weighting}
\end{itemize}
\vspace{-2mm}
Together with whether to include the minimum-phase or inter-aural cross-correlation features, we have 36 configurations in total.
Our implementations and evaluation scripts are publicly available on GitHub~\footnote{\href{https://github.com/iamycy/hrtf-ilp}{github.com/iamycy/hrtf-ilp}}.

\subsection{Reconstruction errors of ITDs and aligned HRIRs}
\label{ssec:eval-recon}

First, we evaluated the accuracy of TOAs for HRIR alignment.
Similar to~\cite{porschmann_cross-correlation-based_2023}, we encoded the time-aligned HRIRs ($h_{i}[t + \tau_{i}]$) and ITDs ($\tau_{l, i} - \tau_{r, i}$) into real SH basis functions.
We set the SH order to four for both features, sufficient to capture the variations in the aligned HRIRs (see Fig.~\ref{fig:sonicom-noise}). 
We measured the \textbf{ITD distortion} ($\frac{1}{N}\sum_{i=0}^N |\text{ITD}_i - \hat{\text{ITD}}_i|$) and \textbf{log-spectral distance (LSD)} between the original and the decoded ITDs and HRIRs average across both ears.
The SH order is proportional to its encoding capacity, and the reconstruction metrics implicitly tell us the amount of higher-order SHs (non-smooth) that are not encoded.
Using the $L^2$ criterion is more likely to result in \emph{over-smoothing} TOAs, which is good for ITD reconstruction but worse for accurate HRIR alignment reflected by LSD. 

Table~\ref{tab:method_compare} verifies our hypothesis, where \textbf{LS} generally achieves the lowest ITD distortion while having higher LSD, meaning its TOAs are too smooth to be accurate.
On the contrary, the TOAs we got from \textbf{EDGY} and \textbf{SIMP} align HRIRs better.
\textbf{EDGY} and \textbf{SIMP} achieve very similar but not identical performances, with \textbf{EDGY} having a slight advantage in ITDs reconstruction.
After inspection, we found that 30\% to 40\% of the time, they arrive at the same solutions, and they have very few differences the rest of the time, indicating multiple optimal solutions exist in our ILP problem.
Regarding the weighting schemes, \textbf{EXP} performs the best in both metrics and all the datasets on average.
\textbf{CORR} gives a slight advantage over uniform weights.

Table~\ref{tab:best_config} shows that 6 out of 7 datasets achieve the lowest LSD with the $L^1$ criterion.
The best weighting scheme varies among datasets, and in contrast to Table~\ref{tab:method_compare}, only two datasets use \textbf{EXP} for the best performance.
Besides ARI and RIEC, the extra correlation features do not help for most datasets.
These findings show that our proposed ILP method generally performs the best.

\subsection{Robustness to noise}

\begin{figure*}[h!]
    \centering
    \includegraphics[width=1\textwidth]{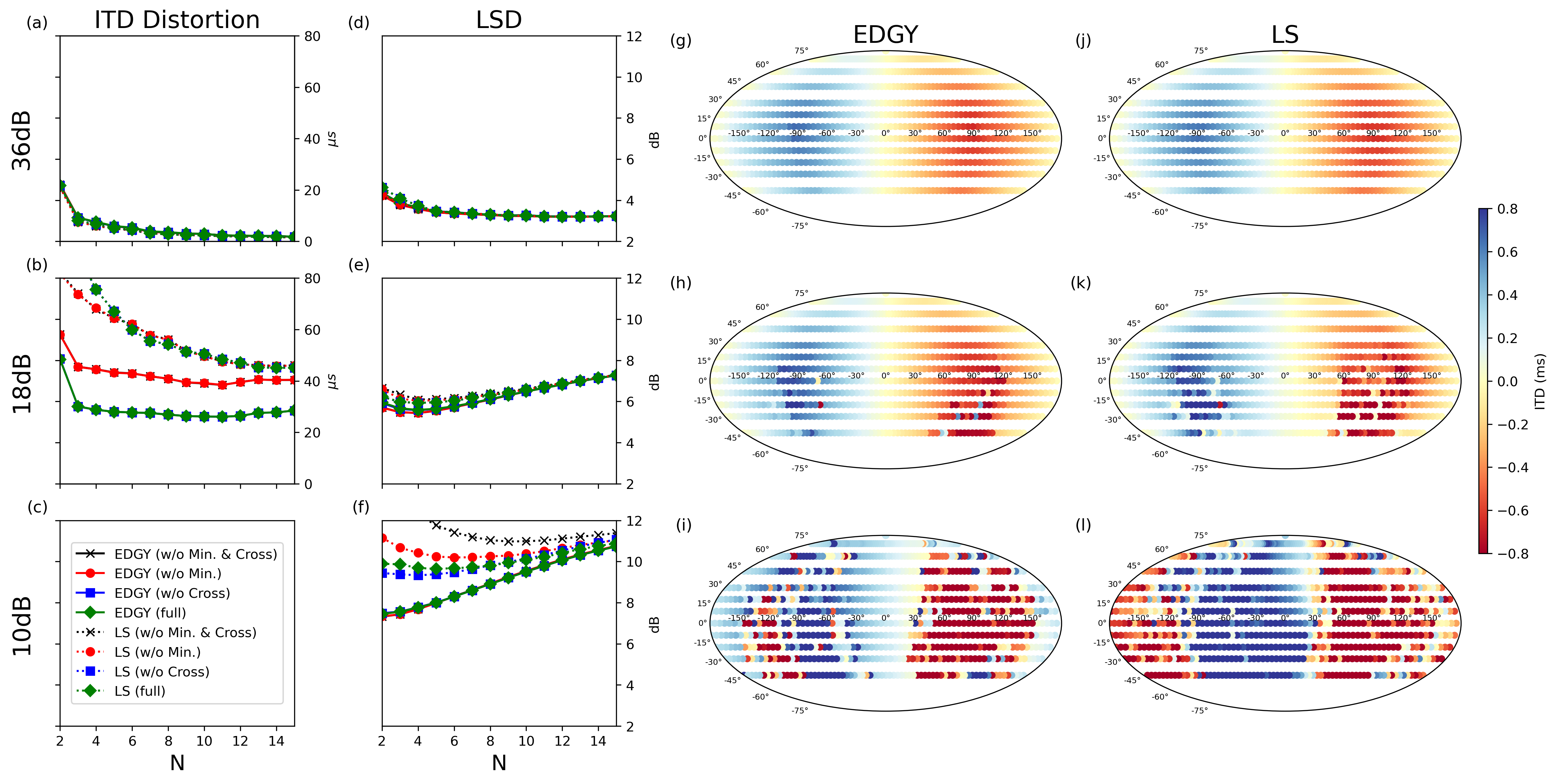}
    \caption{
        Noise robustness experiment on the SONICOM HRTF.
        The LSD (d-f) is calculated between the measured (clean) and the reconstructed noisy HRTFs.
        $N$ is the SH order.
        (g-l): visualisations of ITDs using all the correlation features (full).
        Each row shares the same noise SNR.
        Mollweide projection is used to plot the hemisphere, and each dot is a sampled direction.
        The ITDs are clipped to $\pm 0.8$ ms.}
    \label{fig:sonicom-noise}
\end{figure*}

\begin{figure*}[h!]
    \centering
    \includegraphics[width=1\textwidth]{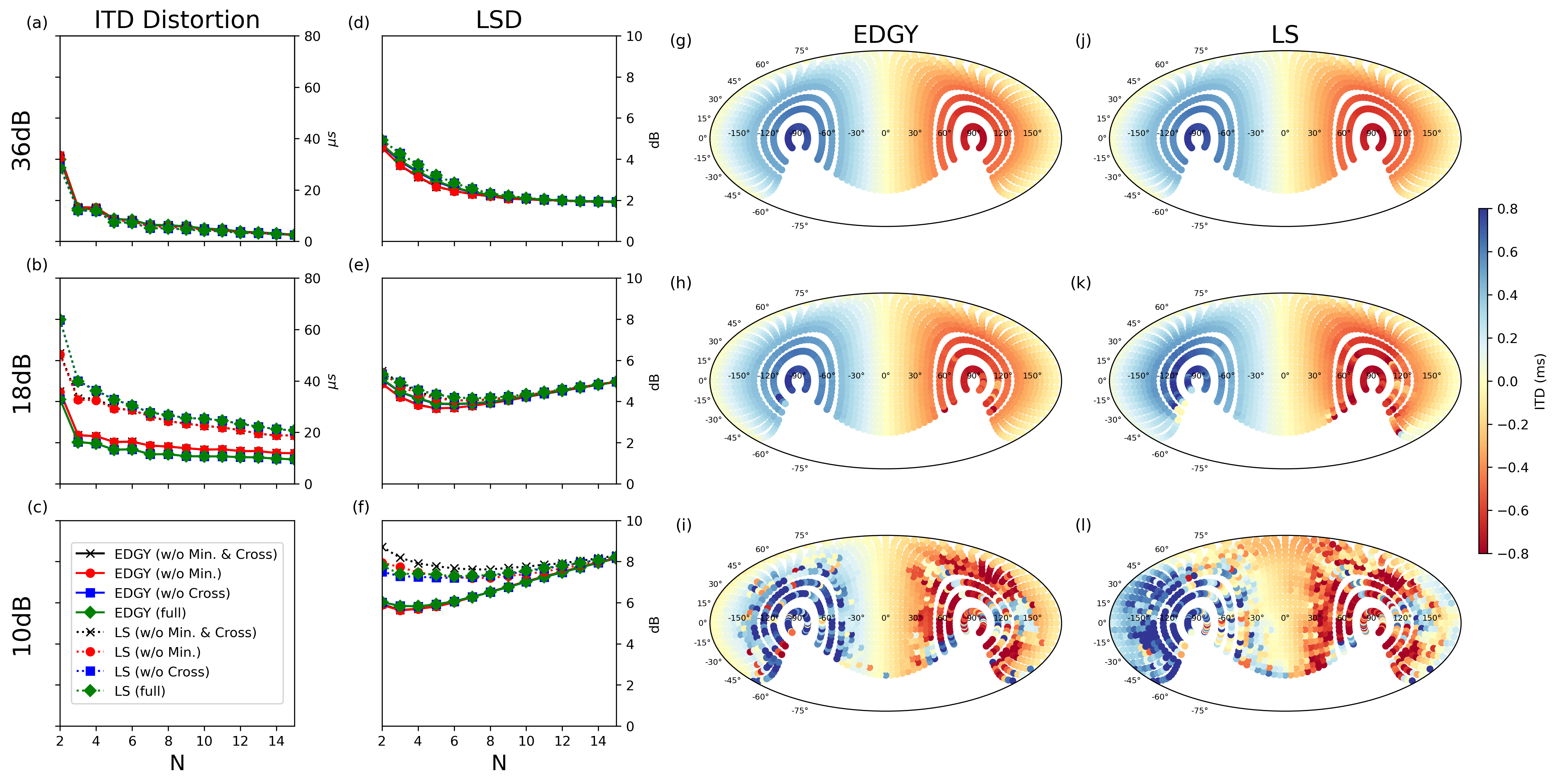}
    \caption{
        Noise robustness experiment on the CIPIC HRTF.
        The details for each subplot are the same as Fig.~\ref{fig:sonicom-noise}.
    }
    \label{fig:cipic-noise}
\end{figure*}

We added random white noise with different SNRs to the SONICOM and CIPIC HRTFs from Table~\ref{tab:datasets}.
Similar to~\citet{gardner1995hrtf}, we calculated the measurement SNR by selecting the HRIRs at the front (0$\degree$ azimuth, 90$\degree$ colatitude) and treating the top 10\% and last 10\% audio samples as signals and noises, resulting in around 75 dB for SONICOM and 60 dB for CIPIC.
We then used these values to simulate different noisy conditions.
\textbf{EXP} weighting is used in this experiment.
We pick \textbf{EDGY} as opposed to \textbf{SIMP} due to its lower average runtime (Table~\ref{tab:datasets}).

From Fig.~\ref{fig:sonicom-noise} and Fig.~\ref{fig:cipic-noise}, we see that with high SNR, both \textbf{EDGY} and \textbf{LS} behave similarly, and there is no difference in including extra features or not.
When SNR decreases ($\leq 18$ dB), \textbf{LS} produces more spikes in ITDs than \textbf{EDGY}, especially (l) from both figures, the wrong estimations spreading the whole sphere.
This effect can also be seen from ITD distortion (b) where \textbf{LS} has higher reconstruction errors.

As \textbf{EDGY} is less sensitive to noise, it aligns the HRIRs better in low SNR, and the effect gets more prominent with lower SH order (e-f).
We also see differences in the use of correlation features.
For \textbf{EDGY}, excluding minimum-phase correlations aligns HRIRs slightly better.
For \textbf{LS}, including extra features always outperforms just using intra-aural correlation.
However, \textbf{LS (full)} is not the best performant, but \textbf{LS (w/o Cross)} is, implying there is a negative effect on minimum-phase correlation from cross-correlation features in low SNR conditions.

\subsection{Reconstruction errors of phase delays}

\begin{figure}[h!]
    \centering
    \includegraphics[width=1\columnwidth]{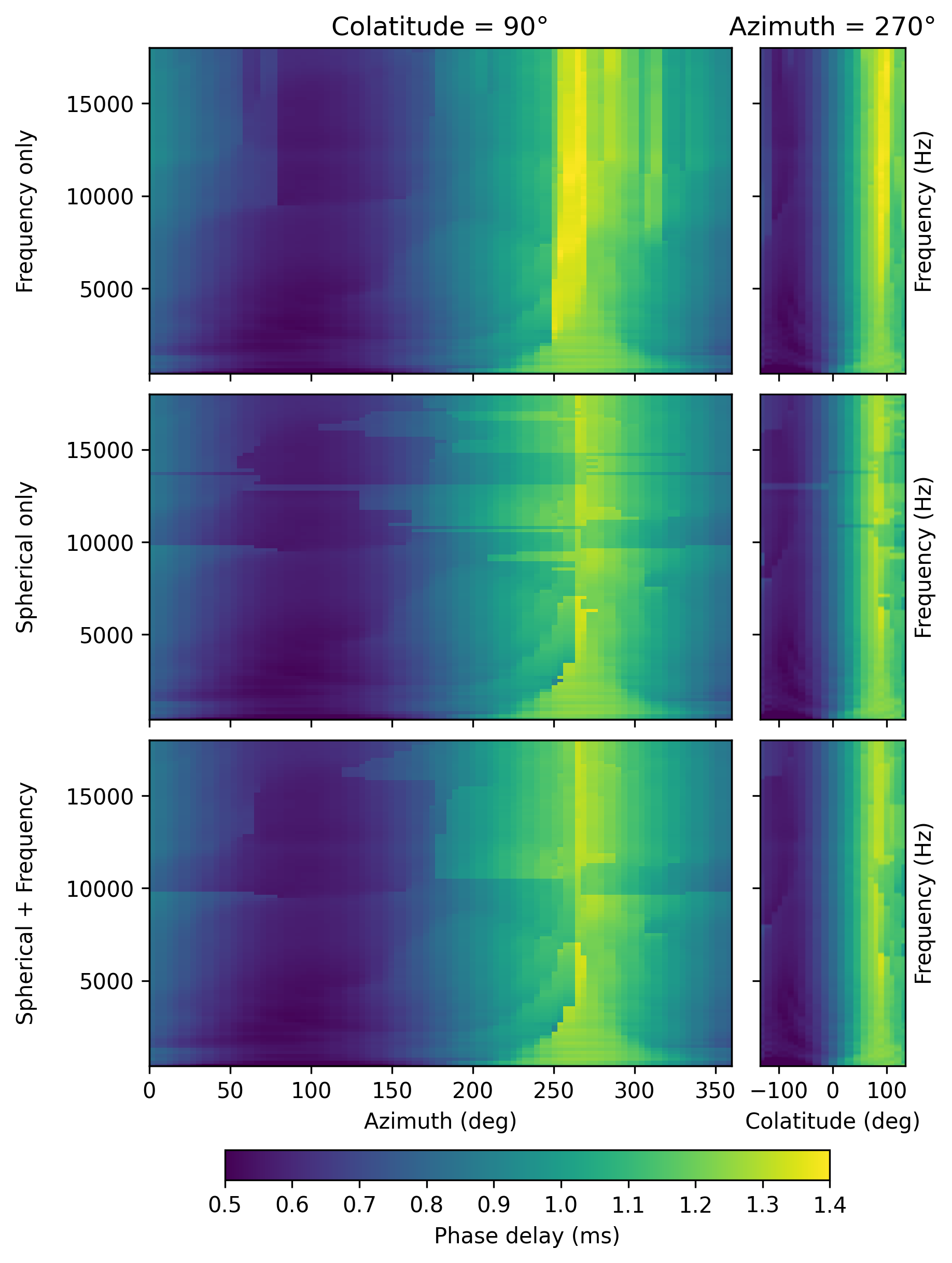}
    \caption{Unwrapped phase delay of SONICOM HRTFs. The left column is the horizontal plane, and the right is the frontal plane. Rows from top to bottom: Sec.~\ref{ssec:freq-pu}, Sec.~\ref{ssec:sph-pu}, and Sec.~\ref{ssec:sph-freq-pu}.}
    \label{fig:sonicom-unwrap}
\end{figure}

\begin{figure}[h!]
    \centering
    \includegraphics[width=1\columnwidth]{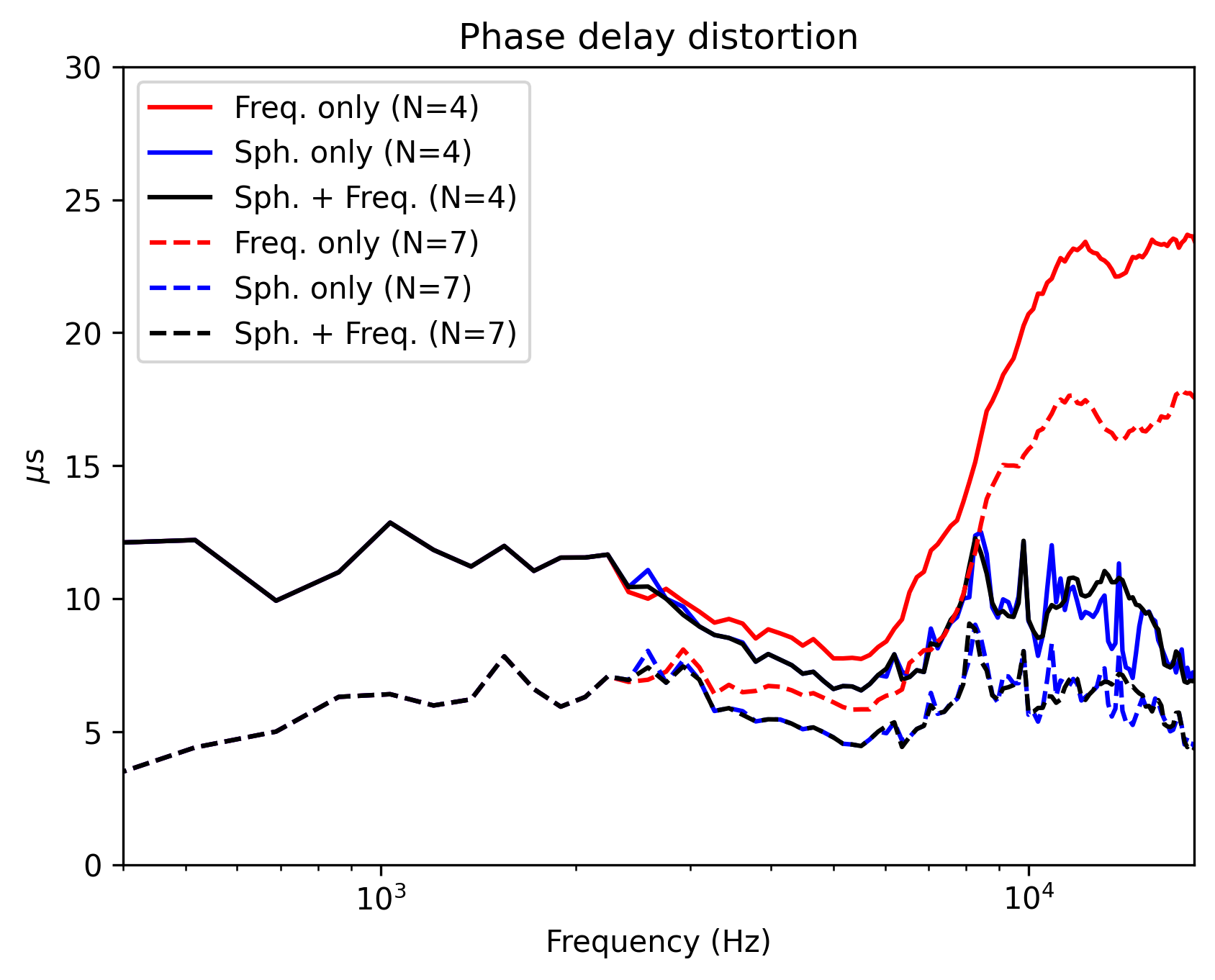}
    \caption{Phase delay distortion as a function of frequency with different unwrapping methods and SH order $N$.}
    \label{fig:phase-dist}
\end{figure}

We used the left ear from SONICOM HRTFs for PU.
Due to the lack of public implementation of~\citet{zaar2011phase}, we simulated spherical unwrapping using ILP by solving each frequency separately and then naively concatenating the results with the number of phase jumps at the minimum between frequencies.

We convert the result to phase delay $-\frac{\phi[f]}{\omega_f}$ and show them in Fig.~\ref{fig:sonicom-unwrap}.
Frequency unwrapping introduces phase jumps that create discontinuities along the spherical (horizontal in the figure) directions; on the other hand, spherical unwrapping has a lot of discontinuities along the frequency axis (vertical in the figure).
Our method behaves somewhere between the two, with discontinuity boundaries crossing both dimensions.

To see the smoothness of unwrapped phases, we ran the same experiment from Sec.~\ref{ssec:eval-recon} on phase delays for each frequency and calculated reconstruction errors the same way as for ITDs.
The result is shown in Fig.~\ref{fig:phase-dist}.
Once the frequency is above 6 kHz, the error starts climbing drastically for frequency unwrapping.
Our method performs comparable to spherical unwrapping, but the error varies smoother for frequency > 10 kHz.
This phenomenon implicitly indicates that the unwrapped phase sphere at each frequency is similar to its nearby frequencies, thus having similar distortion.

\section{Summary}

In this work, we unify the problem of TOA estimation and PU of HRTFs with a graph-based data structure.
By changing the data unit of graph edges, we can solve both problems with the same ILP equation.

For TOA estimation, the proposed method is a variant of the previous SOTA with an $L^1$ criterion instead.
We show the $L^1$ criterion makes more accurate HRIR alignment on seven HRTFs and is more robust than the previous SOTA in extremely noisy conditions on two selected HRTFs.
We also demonstrate the flexibility of our method to include other time difference measurements, providing extra guidance and constraints to enhance performance in noisy conditions.

For PU, our method is the first to utilise all available phase differences and unwrap them jointly per ear.
We show that the proposed method has less spatial discontinuity than naive unwrapping and less frequency discontinuity than pure spherical unwrapping.
Our method can be improved further with custom weights or more advanced graph-based PU, such as the branch cut method, which is left for future work.

\section{Acknowledgements}
The first author is a research student at the UKRI CDT Training in Artificial Intelligence and Music, supported jointly by UKRI [grant number EP/S022694/1] and Queen Mary University of London.

\bibliographystyle{abbrvnat}

\bibliography{clean_ref}

\begin{thebibliography}{20}
\providecommand{\natexlab}[1]{#1}
\providecommand{\url}[1]{\texttt{#1}}
\expandafter\ifx\csname urlstyle\endcsname\relax
  \providecommand{\doi}[1]{doi: #1}\else
  \providecommand{\doi}{doi: \begingroup \urlstyle{rm}\Url}\fi

\bibitem[Arend et~al.(2023)Arend, Pörschmann, Weinzierl, and Brinkmann]{arend2023mag}
J.~M. Arend, C.~Pörschmann, S.~Weinzierl, and F.~Brinkmann.
\newblock Magnitude-corrected and time-aligned interpolation of head-related transfer functions.
\newblock \emph{IEEE/ACM TASLP}, 31:\penalty0 3783--3799, 2023.

\bibitem[Ben-Hur et~al.(2019)Ben-Hur, Alon, Mehra, and Rafaely]{ben-hur_efficient_2019}
Z.~Ben-Hur, D.~L. Alon, R.~Mehra, and B.~Rafaely.
\newblock Efficient {Representation} and {Sparse} {Sampling} of {Head}-{Related} {Transfer} {Functions} {Using} {Phase}-{Correction} {Based} on {Ear} {Alignment}.
\newblock \emph{IEEE/ACM TASLP}, 27\penalty0 (12):\penalty0 2249--2262, Dec. 2019.

\bibitem[Brinkmann and Weinzierl(2018)]{brinkmann2018comparison}
F.~Brinkmann and S.~Weinzierl.
\newblock Comparison of head-related transfer functions pre-processing techniques for spherical harmonics decomposition.
\newblock In \emph{AES International Conference on Audio for Virtual and Augmented Reality}, 2018.

\bibitem[Costantini(1998)]{costantini_novel_1998}
M.~Costantini.
\newblock A novel phase unwrapping method based on network programming.
\newblock \emph{IEEE Transactions on Geoscience and Remote Sensing}, 36\penalty0 (3):\penalty0 813--821, May 1998.

\bibitem[Duraiswami et~al.(2004)Duraiswami, Zotkin, and Gumerov]{duraiswami_interpolation_nodate}
R.~Duraiswami, D.~Zotkin, and N.~Gumerov.
\newblock Interpolation and range extrapolation of hrtfs.
\newblock In \emph{ICASSP}, volume~4, pages iv--iv, 2004.

\bibitem[Engel et~al.(2023)Engel, Daugintis, Vicente, Hogg, Pauwels, Tournier, and Picinali]{engel_sonicom_2023}
I.~Engel, R.~Daugintis, T.~Vicente, A.~O.~T. Hogg, J.~Pauwels, A.~J. Tournier, and L.~Picinali.
\newblock The {SONICOM} {HRTF} {Dataset}.
\newblock \emph{Journal of the Audio Engineering Society}, 71\penalty0 (5):\penalty0 241--253, May 2023.

\bibitem[Evans et~al.(1998)Evans, Angus, and Tew]{evans1998analyzing}
M.~J. Evans, J.~A. Angus, and A.~I. Tew.
\newblock Analyzing head-related transfer function measurements using surface spherical harmonics.
\newblock \emph{The Journal of the Acoustical Society of America}, 104\penalty0 (4):\penalty0 2400--2411, 1998.

\bibitem[Gamper(2013)]{gamper2013selection}
H.~Gamper.
\newblock Selection and interpolation of head-related transfer functions for rendering moving virtual sound sources.
\newblock In \emph{DAFx}, 2013.

\bibitem[Gardner and Martin(1995)]{gardner1995hrtf}
W.~G. Gardner and K.~D. Martin.
\newblock Hrtf measurements of a kemar.
\newblock \emph{The Journal of the Acoustical Society of America}, 97\penalty0 (6):\penalty0 3907--3908, 1995.

\bibitem[Itoh(1982)]{Itoh:82}
K.~Itoh.
\newblock Analysis of the phase unwrapping algorithm.
\newblock \emph{Appl. Opt.}, 21\penalty0 (14):\penalty0 2470--2470, Jul 1982.

\bibitem[Kistler and Wightman(1992)]{kistler1992model}
D.~J. Kistler and F.~L. Wightman.
\newblock A model of head-related transfer functions based on principal components analysis and minimum-phase reconstruction.
\newblock \emph{The Journal of the Acoustical Society of America}, 91\penalty0 (3):\penalty0 1637--1647, 1992.

\bibitem[Ma et~al.(2022)Ma, Jiang, Khoshmanesh, and Cheng]{ma_time_2022}
Z.-F. Ma, M.~Jiang, M.~Khoshmanesh, and X.~Cheng.
\newblock Time {Series} {Phase} {Unwrapping} {Based} on {Graph} {Theory} and {Compressed} {Sensing}.
\newblock \emph{IEEE Transactions on Geoscience and Remote Sensing}, 60:\penalty0 1--12, 2022.

\bibitem[Murphy(2022)]{pml1Book}
K.~P. Murphy.
\newblock \emph{Probabilistic Machine Learning: An introduction}.
\newblock MIT Press, 2022.

\bibitem[Nam et~al.(2008)Nam, Abel, and Smith~III]{nam2008method}
J.~Nam, J.~S. Abel, and J.~O. Smith~III.
\newblock A method for estimating interaural time difference for binaural synthesis.
\newblock In \emph{Audio Engineering Society Convention 125}, 2008.

\bibitem[P{\"o}rschmann et~al.(2023)P{\"o}rschmann, L{\"u}beck, and Arend]{porschmann_cross-correlation-based_2023}
C.~P{\"o}rschmann, T.~L{\"u}beck, and J.~M. Arend.
\newblock Cross-correlation-based time-aligned interpolation of head-related impulse responses.
\newblock In \emph{AES International Conference on Spatial and Immersive Audio}, 2023.

\bibitem[Pörschmann et~al.(2019)Pörschmann, Arend, and Brinkmann]{arend_spatial_nodate}
C.~Pörschmann, J.~M. Arend, and F.~Brinkmann.
\newblock Directional equalization of sparse head-related transfer function sets for spatial upsampling.
\newblock \emph{IEEE/ACM TASLP}, 27\penalty0 (6):\penalty0 1060--1071, 2019.

\bibitem[Reijniers et~al.(2023)Reijniers, Partoens, and Peremans]{reijniers_noise-resistant_2023}
J.~Reijniers, B.~Partoens, and H.~Peremans.
\newblock Noise-resistant correlation-based alignment of head-related transfer functions for high-fidelity spherical harmonics representation.
\newblock In \emph{AES International Conference on Spatial and Immersive Audio}, 2023.

\bibitem[Shanker and Zebker(2010)]{shanker_edgelist_2010}
A.~P. Shanker and H.~Zebker.
\newblock Edgelist phase unwrapping algorithm for time series {InSAR} analysis.
\newblock \emph{JOSA A}, 27\penalty0 (3):\penalty0 605--612, Mar. 2010.

\bibitem[Wang et~al.(2009)Wang, Yin, and Chen]{wang_head-related_2009}
L.~Wang, F.~Yin, and Z.~Chen.
\newblock Head-related transfer function interpolation through multivariate polynomial fitting of principal component weights.
\newblock \emph{Acoustical Science and Technology}, 30\penalty0 (6):\penalty0 395--403, 2009.

\bibitem[Zaar(2011)]{zaar2011phase}
J.~Zaar.
\newblock Phase unwrapping for spherical interpolation of head-related transfer functions.
\newblock \emph{M. thesis, IEM, Univ. of Music and Performing Arts Graz}, 2011.

\end{thebibliography}

\end{document}